\documentstyle[preprint,revtex]{aps}

\begin{document}
\begin{title}
Model for asymptotic $D$-state parameters of light nuclei:\\
Application to $^4He$
\end{title}
\author{Sadhan K. Adhikari}
\begin{instit}
Instituto de F\'\i sica Te\'orica, Universidade Estadual Paulista \\
01405-000 S\~{a}o Paulo, S\~{a}o Paulo, Brasil
\end{instit}
\moreauthors{ T. Frederico}
\begin{instit}
 Instituto de Estudos Avan\c cados,
  Centro T\'ecnico Aeroespacial \\
12231--970    S\~ao Jos\'e dos Campos,  S\~ao Paulo, Brasil\\
\end{instit}
\moreauthors{ I. D. Goldman}
\begin{instit}
 Departamento de F\'\i sica Experimental,
 Universidade de S\~ao Paulo\\
  20516 S\~ao Paulo, S\~ao Paulo, Brasil
\end{instit}
\moreauthors{  S. Shelly Sharma}
\begin{instit}
 Departamento de F\'\i sica, Universidade Estadual
 de Londrina\\
86020 Londrina, Parana, Brasil
\end{instit}

\begin{abstract}
A simple method for calculating the asymptotic $D$ state
observables  for light nuclei is suggested.
 The method exploits the dominant clusters of
the light nuclei. The  method is applied to calculate the $^4He$
asymptotic $D$ to $S$ normalization ratio $\rho^\alpha$ and the closely
related $D$ state parameter $D_2^\alpha$. The study predicts a correlation
between   $D_2^\alpha$ and $B_\alpha$, and between $\rho^\alpha$
and $B_\alpha$,  where $B_\alpha$ is the binding energy of $^4He$. The
present study yields $\rho^\alpha \simeq -0.14$ and  $D_2^\alpha
\simeq -0.12$ fm$^2$ consistent with the correct experimental $\eta^D$ and the
binding energies
of the deuteron, triton, and the $\alpha$ particle,
where $\eta^d$ is the deuteron $D$ state to $S$ state
normalization ratio.

\end{abstract}


\section{INTRODUCTION}

The role of the deuteron asymptotic $D$ to $S$ normalization ratio $\eta^d$
has been emphasized recently in making a theoretical estimate of the triton
asymptotic $D$ to $S$ normalization ratio $\eta^t$\cite{Fred1,Fred2}.
There has been considerable interest for theoretical and experimental
determination of the asymptotic $D$ to $S$ normalization ratio  of
light nuclei ever since Amado suggested that this ratio should be given the
``experimental" status of a single quantity to measure the $D$ state of
light nuclei \cite{Amado}. In this paper we generalize certain ideas used
successfully in the two- and  three-nucleon systems in order to formulate
a model for the asymptotic $D$ to $S$ normalization ratio of light nuclei.
We  apply these ideas to the study of the asymptotic $D$ to $S$
normalization ratio, $\rho^{\alpha}$, and the $D$ state parameter,
$D_2^{\alpha}$, of $^4He$.

Though a realistic numerical study of the asymptotic $D$ to $S$
normalization ratios of $^2H$, $^3H$, and $^3He$  is completely under
control\cite{Girar,Fria1,Ishik,Eric,Friar2,Rodn}, the same can not be
affirmed in the case of other light nuclei. Even in the case of $^4He$,
such a task, employing the Faddeev-Yakubovskii dynamical equations, is a
formidable, but feasible, one.  This is why approximate methods are called
for. As the nucleon-nucleon tensor force plays
a crucial and fundamental role in the formation of the $D$ state of light
nuclei\cite{Fred1,Fred2,Eric}, it is interesting to ask what are the
dominant many body mechanisms that originate the $D$-state.  The present
study is aimed to shed light on the above questions.

In the case of
 the $D$ state of the deuteron, exploiting the weak (perturbative) nature
of the $D$ state, Ericson and Rosa Clot\cite{Eric}  have demonstrated that
the essential ingredients of the asymptotic $D$ to $S$ normalization ratio
$\eta ^d$  are the long-range one-pion-exchange tail of the
nucleon-nucleon interaction, the binding energy, $B_d$, and the $S$-state
asymptotic normalization parameter (ANP), $C^d_S$, of the deuteron.
In the case of $^3H$
we have seen that the long-range one-nucleon-exchange tail of the
nucleon-deuteron interaction plays a crucial role in the formation of the
trinucleon $D$ state\cite{Fred1,Fred2}. We have demonstrated that
all realistic nucleon-nucleon potentials  will virtually yield
the same value of $\eta ^t$ provided that they also yield the same values
for the $S$-state ANP  and $\eta ^d$ of the  deuteron and binding energies of
$^2H$ and $^3H$\cite{Fred1,Fred2}.

The purpose of the present study is to identify the dominant mechanisms for the
formation of the $D$ state in more complex situations. We do not
consider the full dynamical problem for our purpose, but, rather, a cluster
model exploiting the relevant long-range part of the cluster-cluster
interaction  supposed to be responsible for the formation of the relevant
$D$ state.  The $\alpha$ particle or $^4He$ has a very important role
in nuclear physics and a study of its structure deserves a special attention.
One important aspect of its bound state is its $D$ state admixture for the
$^4He\rightarrow\,2\, { ^2}H$ channel.  There have been a lot of theoretical
and experimental activities for measuring the asymptotic $D$ state to $S$ state
normalization ratio $\rho^{\alpha}$ for this channel.
\cite{Weller1,SX0,Weller2,Santos1,Santos2,Karp,Tost1,Tost2,Merz,Schia,Koonin}
In this paper we study the $D$ state of $^4He$ and
make a model independent estimate
of $\rho^{\alpha}$  and the closely related parameter $D_2^{\alpha}$.

All the observables directly sensitive to the tensor force
of the nucleon-nucleon interaction, such as the deuteron quadrupole moment,
$Q^d$, and $\eta ^t$ etc., have been found  to be correlated in numerical
calculations with $C^d_S$ through the relation\cite{Fred1,Fred2,Eric,Adhik1}
\begin{equation}
\frac{{\cal O}}{\eta ^d} \sim ( C^d_S)^2 f,
\label{1}
\end{equation}
where ${\cal O}$ stands for $Q^d$, $\eta^t$, or  the usual $D$ state
parameter, $D^t_2$, for the triton. The function $f$ depends on the relevant
binding energies, e.g., the binding energy of the deuteron in the case of
$Q^d$, and the binding energies of the deuteron and triton in the case of
$\eta^t$ and $D^t_2$, while other low-energy on-shell nucleon-nucleon
observables are held fixed. If correlation (\ref{1}) were exact, no new
information about the nucleon-nucleon interaction  could be obtained from
the study of $Q^d$, $\eta ^t$, or $D^t_2$, which is not implicit in the
values of $B_d$, $B_t$, $C_S^d$, and $\eta ^d$.\cite{Fred1}
However, this correlation
is approximate and information about the nucleon-nucleon tensor
interaction might be obtained from a study of these parameters from a
breakdown of these correlations.  In order that such informations could be
extracted,  however, one should require  precise experimental measurements of
these observables.\cite{Fred1,Fred2}

In this paper we shall be interested to see if  correlation (1)
extrapolates to the case of other light nuclei, specifically, to the case
of $^4He$.  We provide a perturbative solution of the problem, which presents
a good description of  the $D$-state. We find that
in order to reproduce the correct $D$-state parameters of $^4He$, the
minimum ingradients required of a model are the correct low-energy
deuteron properties including $C_S^d$ and $\eta ^d$ and the triton and
$^4He$ binding energies, $B_t$ and $B_{\alpha}$.

The model also provides  the essential behavior of
$D_2^{\alpha}$ and $\rho^\alpha$ as function of the binding energy,
$B_{\alpha}$, of the $\alpha$ particle for  fixed $B_d$, $B_t$, and
$\eta^d$. Consistent with the experimental $B_d$, $B_t$, $B_\alpha$,
and $\eta^d$ we find $\rho^\alpha = -$ 0.14, and $D_2^\alpha = -$0.12
fm$^2$. The model also predicts an approximate  linear correlation between
$D_2^{\alpha} $ ($\rho ^{\alpha}$) and $B_{\alpha}$ for  fixed
$B_d$, $B_t$, and $\eta^d$ to be
 verified in realistic dynamical four-nucleon calculations.

The model for the formation of the $D$ state is given in Sec. II.
 In Sec. III we present
relevant notations for our future development of the $D$ state. In Sec. IV
the analytic model for the $D$ state of $^4He$ is presented. Section V deals
with the numerical investigation of our model. Finally, in Sec. VI
 brief summary and discussion are presented.

\section{THE MODEL}

 As the exact
dynamical studies of the $D$ state for the light nuclear systems
employing the connected kernel Faddeev-Yakubovskii equations are usually
performed in the momentum space, we present our model in the
momentum space in terms of the Green functions or propagators.

Figures 1 and 2 represent a coupled set of dynamical equations between clusters
valid for $ ^2H$ and $ ^3H$, respectively. In the case of the deuteron  the
dashed line denotes the  exchanged meson.
 In the case of the triton the exchanged particle is a nucleon and the double
 line denotes a deuteron. In both cases a single line denotes a nucleon.
 In the case of  the deuteron these equations are essentially the
homogeneous version of the momentum space Lippmann-Schwinger equations
for the nucleon-nucleon system,
which couples the $S$ and the $D$ states of the deuteron. Explicitly, these
equations are written as
\begin{equation}
g_0 =   V_{00} G_0 g_0 + V_{02}G_0 g_2,
\label{2}
\end{equation}
\begin{equation}
g_2 =   V_{20} G_0 g_0 +V_{22}G_0 g_2,
\label{3}
\end{equation}
where $g_l=  V \vert \phi_l \rangle$ ($l=0,2$)
represent the relevant form factors for the two states denoted by
the two-body bound state wavefunction $\phi_l,$
$G_0$ is the free Green function for propagation, and $V$'s are the
relevant potential elements between the $S$ and the $D$ states.
 Figure  1(b) gives the two ways of forming the $D$ state at infinity:
(a) in the first term on the right-hand side (rhs), the deuteron breaks up
first into two nucleons in
the $S$ state which gets changed to two nucleons in the $D$ state via the
one-pion-exchange nucleon-nucleon tensor force, (b) in the second term
on the rhs, the deuteron
breaks up first into two nucleons in the $D$ state which
continues the same under the action of the central one-pion-exchange
nucleon-nucleon interaction.  As the $D$  state of the deuteron could be
considered to be a perturbative correction on the $S$ state, in Fig. 1(b)
the first term on the rhs is supposed to dominate, with
the second term providing small correction.  Hence, the essential mechanism
for the formation of the $D$ state in this case
is given by the following equation
\begin{equation}
g_2 =   V_{20} G_0 g_0.
\label{4}
\end{equation}
Given a reasonable $g_0$ and the tensor interaction $V_{20}$, Eq. (\ref{4})
could be utilized for studying various properties of the $D$ state. This
equation should determine the asymptotic $D$ to $S$ ratio of deuteron
$\eta^d$ provided that the model has the correct deuteron binding $B_d$
and the one-pion-exchange tail of the tensor nucleon-nucleon interaction.

In the momentum space representation of Eq. (\ref{4}), at the bound state
energy,  $g$'s have the following structure
\begin{equation}
\langle i \mu \vert g_l \rangle \sim  {C_l^d} ,
\label{J}
\end{equation}
with $\mu = \sqrt{2 m_R B_d}$, $m_R$ being the reduced mass and  $C_l^d$ the
deuteron ANP's for the state of angular momentum $l$. The off-diagonal
tensor potential $V_{20}$ is proportional to $g_{\pi N}^2$, where $g_{\pi
N}$ is the pion-nucleon coupling constant. From Eqs. (\ref{4}) and
(\ref{J}), at the bound state energy one has
\begin{equation}
\eta ^d \sim g_{\pi N}^2 \times Int,
\label{5}
\end{equation}
where $Int$ represents a definite integral determined by the deuteron binding
$B_d$. Hence $\eta ^d$ is mainly determined by the deuteron binding energy
and the pion-nucleon coupling constant\cite{Eric}.

This idea could be readily generalized to more complex situations.  In the
case of the triton $D$-state, Fig. 2 and Eqs. (\ref{2}) and (\ref{3})
are valid. The form-factors $g_l$ are to be interpreted as the
triton-nucleon-deuteron form factors, the Green function $G_0$ represents
the free propagation of the nucleon-deuteron system, and
 the potentials $V_{02}$ and $V_{20}$ are the  Born
approximation to the rearrangement nucleon-deuteron elastic sacttering
amplitudes representing the transition between the relative $S$ and $D$
angular momentum states of the nucleon-deuteron system. For example, for
nucleon-nucleon
separable tensor potential, $V_{02}$ corresponds to the inhomogeneous term
of the Amado model\cite{Adhik1}
for nucleon-deuteron scattering for the transition
between $S$ and $D$ states of the nucleon-deuteron system. The essential
mechanism for the formation of the $D$-state is again given by Eq. (\ref{4}).
Now in the momentum space representation of Eq. (\ref{4}), at the bound state
energy, $g$'s have essentially the structure given by
\begin{equation}
\langle i \mu \vert g_l  \rangle \sim  {C_l^t} ,
\label{6}
\end{equation}
where $C_l^t$ is the ANP of the triton for the angular momentum state $l$.
In Eq. (\ref{4}),  $V_{02}$ connects a relative nucleon-deuteron $S$ state
to a nucleon-deuteron $D$ state in different subclusters via a nucleon
exchange. Hence the amplitude $V_{02}$
 involves two form-factors, one for the deuteron $S$ state and the
other for the deuteron $D$ state. Consequently, at the triton pole the
momentum space version of Eq. (\ref{4}) has the following form:
\begin{equation}
 C_D^t \sim \,C_S^d\, C_D^d\, C_S^t \,\,Int,
 \label{7} \end{equation}
where $Int$ represents the remaining definite integral now expected to be
determined essentially by the deuteron and triton binding energies and other
low-energy nucleon-nucleon observables. Recalling that $\eta ^t \equiv
C_D^t/C_S^t$, with $\eta ^d$ defined similarly, Eq. (\ref{7}) reduces to
Eq. (\ref{1}). Hence, this simple consideration shows that the ratio
$\eta ^t/\eta ^d$ is a universal one satisfying Eq. (\ref{1})
determined essentially by the deuteron and triton binding energies
and the deuteron $S$ wave ANP $C_S^d$.

Next let us consider the example of ${^4He}$, where the two deuterons could
appear asymptotically either in a relative $S$ or a $D$ state. However,
asymptotically the nucleon and the trinucleon could exist only in the
relative $S$ state.  In this
case the lowest scattering thresholds are the nucleon-trinucleon and the
deuteron-deuteron ones.  If we include these two possibilities of breakup
of ${^4He}$, then the principal mechanisms for the formation of the asymptotic
deuteron-deuteron states are given in
 Fig. 3.  We have two  equations of the type shown in
Fig. 3, one for the $S$ state and the other for the $D$ state.
In Fig. 3 the contribution of the last term on the rhs is expected to be
 small.  The virtual breakup of ${^4He}$ first to two  deuterons and their
eventual breakup to four nucleons to form the four-nucleon-exchange
deuteron-deuteron amplitude  as in this term is much less probable at
negative energies than the virtual breakup of ${^4He}$ to a
nucleon and a trinucleon and its eventual transformation to the
deuteron-deuteron cluster as in the first term on the right-hand side of
this equation. For this reason we shall neglect the last term of Fig. 3 in
the present  treatment.  As in the three-nucleon case the amplitudes in
Fig. 3 are the Born approximations to  rearrangement amplitudes between
diffenent subclusters, which connect different angular momentum states,
e.g., $S$ and $D$.

We notice that in the first term of Fig. 3 either of the vertices has
to be a $D$ state so that the passage from $S$ to $D$ state is allowed in this
diagram.  Consequently, at the pole of the ${ ^4He}$ bound state
the momentum space version of Fig. 3 has two contributions corresponding
to the deuteron (triton) vertex on the right hand side being the
$S$-state  and the triton (deuteron) vertex being the $D$-state so that we may
write,
\begin{equation}
C_D^{\alpha \rightarrow dd} \sim \,C_S^{\alpha \rightarrow nt}\, C_S^d
\,C_D^t\,\,  Int_1 +
C_S^{\alpha \rightarrow nt}\, C_D^d\, C_S^t\,\,  Int_2,
\label{9} \end{equation}
 where $Int_1$ and $Int_2$ are two definite integrals.
 The $ ^4He$ asymptotic $D$-state to $S$-state ratio
$\rho^\alpha$ is defined by
\begin{equation}
 \rho^{\alpha} \equiv \frac{C_D^{\alpha \rightarrow dd}}{C_S^{\alpha
  \rightarrow dd}}.
\label{10} \end{equation}
 It is clear that, unlike in the case of triton, $\rho^{\alpha}$ is determined
 by two independent terms. Physically, it means that there are two mechanisms
that construct the $D$ state ANP of $^4He$.
 Now recalling the empirical relation $\eta^t \equiv {C_D^t}/{C_S^t}
 \sim ( C_S^d )^2
\eta^d$,  we obtain from Eq. (\ref{9})
\begin{equation}
\frac{\rho^{\alpha}}{{\eta}^d} \sim  \xi_S\,\, Int,
\label{11}\end{equation}
where $\xi_S$ is determined by the $S$-state asymptotic normalizations
${C_S^{\alpha \rightarrow dd}}$, ${C_S^{\alpha \rightarrow nt}}$, $C_S^t$,
$C_S^d$, and  $Int$ represents integrals which are essentially determined
by the binding energies $B_d$, $B_t$ and $B_{\alpha}$. Hence,  in the case
of $^4He$ Eq. (\ref{1}) gets modified to one of the form of Eq. (\ref{11}).

\section{DEFINITIONS AND NOTATIONS}

In this Section we present  notations and definitions which we shall use
for future development.
The asymptotic wavefunction for a two-body bound state, $\phi_l$, (binding
energy $B$) in a potential $V$ is given by
\begin{equation}
\lim_{r \rightarrow \infty } \langle rlj|\phi_{l} \rangle = -\frac
{\sqrt{2 \pi}\, m_R\, e^{- \mu r}}{r} \lim_{q \rightarrow i \mu} \langle
qlj |V|\phi_l \rangle,
\label{12}\end{equation}
where $l$ is the relative orbital angular momentum, $j$ is the total final
spin of the system (the intrinsic spin of the system is not shown) and
  $| qlj \rangle$ is the momentum space
wave function.  The asymptotic normalization
parameter $C_{jl}$ for this state is defined by
\begin{equation}
\lim_{r \rightarrow \infty }{\langle rlj|\phi_{l} \rangle} = \frac
{C_{jl}\, \sqrt{2 \mu}\ e^{- \mu r}}{\sqrt{N}\,r}.
\label{13}\end{equation}
Here $N$ represents the number of ways a particular asymptotic
configuration can be constructed from its constituents in the
same channel. For example, in the  channel, $ { ^{3}}H
\rightarrow  n+{ ^{2}}H $, as we can combine the proton with
either of the two neutrons to form ${ ^{2}}H$, $N=2$. Similarly
in the $ { ^{4}}He \rightarrow 2\,{ ^{2}}H $ channel, the
deuteron can be formed in two different ways and $N=2$. But
in the channels $ { ^{4}}He
\rightarrow \, ^{3}H + ^{1}H$ and $ { ^{4}}He \rightarrow n+\,
^{3}He $, neglecting Coulomb interaction, there are four
different possibilities for constructing the outgoing channel
components, so that $N=4$.

{}From Eqs. (\ref{12}) and (\ref{13}), we obtain
\begin{equation}
C_{jl} =- \frac{m_R\,\sqrt{ \pi N}}{\sqrt
\mu}  \lim_{q \rightarrow i \mu} \langle
qlj |V|\phi_l \rangle .
\label{14} \end{equation}
As the partial wave $t$ matrix  may be expressed as
\begin{equation}
\langle qlj |t|qlj \rangle=\lim_{q \rightarrow i \mu} \frac {|\langle qlj
|V|\phi_l \rangle|^2}{E+B},
\label{15}\end{equation}
the parameter $C_{jl}$ is  related to the residue at the $t$-matrix
pole by
\begin{equation}
\langle qlj |t|qlj \rangle_{Res}\equiv \lim_{q \rightarrow i \mu}
 {|\langle qlj|V|\phi_l \rangle|^2} =\frac{\mu}{\pi N m_R^2}\,C_{jl}^2.
\label{16}\end{equation}
With this definition, in the limit of $\mu\rightarrow 0$,
$C_{j\ell}\rightarrow 1$.\cite{Fred1}

 For the two-particle bound state, the  vertex function $g(q)$
 for  a definite angular momentum $l$ (and $j$) can be written as
\begin{equation}
g_{j\ell} (q^2)  =
- \sqrt{ \frac{\mu}{\pi N m_R^2}}\,C_{jl}\, ( \frac{q}{i \mu})^l
\,{\hat g}(q^2),
\label{17}\end{equation}
where the kinematical factor which takes into account the centrifugal
barrier has been explicitly shown. The function ${\hat g}(q^2)$
essentially provides the momentum dependence of the vertex function. In
the present qualitative study we set ${\hat g}(q^2)=1$  so that we have
\begin{equation}
g_{j\ell} (q^2)  =
- \sqrt{ \frac{\mu}{\pi N m_R^2}}\,C_{jl}\, ( \frac{q}{i \mu})^l\,.
\label{17A}\end{equation}
In Eq. (\ref{17A}) apart
 from a kinematical factor that takes into account the centrifugal
barrier, the form factor is assumed to be independent of the relative
momentum of the two components forming the bound state, consistent with
the  minimal three body model \cite{Fred1}.
In particular, the form factors for the formation of ${ ^4He}$ from
nucleon($N$)$-$triton($t$) channel and from two deuterons($dd$) are
respectively,
\begin{equation}
g_{00}^{Nt} (q) = -\frac {2}{3} \frac{ \sqrt{\mu_{Nt}}}{ \sqrt{ \pi}}
C_{S}^{\alpha \rightarrow Nt},
\label{18}\end{equation}
and
\begin{equation} g_{jl}^{dd} (q) = -\frac{ \sqrt{\mu_{dd}}}{ \sqrt{2
\pi}} C_{jl}^{\alpha \rightarrow dd} (\frac{q}{i
\mu_{dd}} )^l,
\label{19}\end{equation}
with
\begin{equation}
\mu_{Nt}=\sqrt{3(B_{\alpha}-B_t)\over 2}\quad, \quad
\mu_{dd}=\sqrt{2(B_{\alpha} - 2 B_d)},
\label{20}\end{equation}
where $B_{\alpha}$ and $B_d$ are the binding energies of ${ ^{4}}He$ and
${ ^{2}}H$ respectively and we assume $\hbar=m_{nucleon}=1$.

For the case where angular momentum states $S$ and $D$ states are mixed,
the probability amplitude for a given $l$-value is proportional to the
corresponding spherical harmonic $Y_{lm_l}$($\hat  q$). Defining the
spin-angular momentum functions ${\cal Y}_{sljm}(\hat q)$ as
\begin{equation}
  {\cal Y}_{lsjm}(\hat q)=(Y_l(\hat q) \bigotimes \psi_s)_{jm}=
  \sum_{m_s,m_l} C^{lsj}_{m_l m_s m} Y_{lm_l}(\hat q) \psi_{s m_s},
\label{21}\end{equation}
where $\psi_s$ is the spin state of the system,
$\otimes$ denotes angular momentum coupling,
the vertex functions in the minimal model for $t \rightarrow Nd$ and $ d
\rightarrow NN$ vertices take the form
\begin{equation}
{\bf g}_{j={1\over2}, m}^{Nd} (\vec p_1) = -\frac {3}{2}
\sqrt{\frac {\mu_{Nd}}{2 \pi}}
C_S^{t} \left[ {\cal Y}_{0{1 \over 2} {1 \over 2}m}(\hat p_1)
-\frac{\eta^t p_1^2}{\mu_{Nd}^2} {\cal Y}_{2{3 \over 2}{1 \over
2}m}(\hat p_1) \right]
\label{22} \end{equation}
\begin{equation}
{\bf g}_{j=1, m}^{NN} (\vec p_3) = - \sqrt { \frac{4
\mu_{NN}}{\pi}}C_S^{d}\left[{\cal Y}_{011m}(\hat p_3)-\frac {\eta^d
p_3^2}{\mu_{NN}^2}{\cal Y}_{211m}(\hat p_3)\right]
\label{23}\end{equation}

Here $C_S^{d}$ and $C_S^{t}$ are the asymptotic
normalization parameters for the $S$ state of the deuteron and triton
respectively, whereas $\eta^d$ and $\eta^t$ are the ratios of corresponding
$D$-state ANP's with the $S$-state ANP's. The relative momentum of the
nucleon with respect to the deuteron is $\vec{p}_1$, whereas the relative
momentum of the two nucleon system is $\vec{p}_3$.

\section{D-STATE PARAMETERS OF $^4He$}

Having defined the relevant vertex functions, we can write down the equation
for constructing the $S$ and $D$ states of $^4He$ in the $^4He \rightarrow
2\,  ^2H$ channel. The first line of
Fig. 3 represents diagrammatically the present
 model for the formation of the asymptotic $S$ and $D$ states of $^4He$.
Using the notation of previous  Sec. the explicit partial wave
form of the present model could be written down as:
\begin{eqnarray}
g_{0l_3}^{dd} (q_3) & = &\int_0^{\infty} dq_1 q_1^2
\int \int d{\Omega_{q1}}d{\Omega_{q3}}
\sum_{L1,L3}\left( \left( {\bf\hat g}_{1L_3}^{NN}({\vec{p}}_3) \bigotimes
{\chi_1^d}\right)^\ast_{l_3} \bigotimes Y_{l_3}^\ast
(\hat{q}_3)\right)_{00} \nonumber \\ & \times  &
\frac{1}{B_d-B_{\alpha}-q_1^2-\frac{3}{4}q_3^2-\vec{q}_1 \cdot \vec{q}_3}
\left(\left( {\bf\hat g}_{{1\over2}L_1}^{Nd}(\vec{p}_1) \bigotimes
\chi^N_{1 \over2}
\right)_{0} \bigotimes Y_{0}(\hat{q}_1) \right)_{00} \nonumber \\ & \times  &
\frac{1}{B_t - B_\alpha-{2 \over3} q_1^2}\,{g}_{00}^{Nt} (q_1)  ,
\label{24}
\end{eqnarray}
where $\vec{p}_1 = -({2 \over3} \vec{q}_1+\vec{q}_3 )$ is the relative
momentum between the the exchanged nucleon $2$ and the structureless deutron
3, $\vec{p}_3 = (\vec{q}_1 +
{{\vec{q}_3}\over2})$ is the relative momentum between the spectator
nucleon $1$ and nucleon $2$, $q_1$ is the momentum carried by nucleon 1 and
$q_3$ is the momentum carried by the structureless
deutron of momentum $q_3$. The indices $1, 2$ refer to the spectator nucleon,
the exchanged nucleon,
and $3$ refers to the structureless deuteron. Here
$l_3$ is the angular momentum state of $^4He$; $l_3 =0 $ (2) corresponds to
the $S$ ($D$) states of $^4He$.
$L_3$ is the relative
angular momentum of the two nucleons forming the deuteron of momentum $-q_3$,
$L_1$ is the relative angular momentum of the structureless
deuteron of momentum $q_3$ and the
nucleon 2 forming the triton of momentum $-q_1$, $\chi^N_{1 \over2}$ is the
spin state of nucleon  1 with momentum $q_1$, and $\chi^d_1$ is the
spin state of the structureless deuteron  with momentum $q_3$.
We dropped the index $m$ of the form-factors
at $NN$, $Nd$ and $Nt$ vertices, because of the angular momentum coupling
notation employed.
Here ${\bf\hat g}_{jL}$ is the $L$ component of the vertex defined
in Eqs. (\ref{22}) and (\ref{23}).   For example,
\begin{equation}
{\bf \hat g}_{1 L_3}^{NN} (\vec p_3) = - \sqrt { \frac{4
\mu_{NN}}{\pi}}C_{L_3}^{d}{\cal Y}_{L_311m}(\hat p_3),
\label{223} \end{equation}
where $C_{L_3}^{d} = C_S^d$  $(-C_S^d\eta^dp_3^2/{\mu_{NN}^2})$ for
$L_3 = 0$ (2).

The rhs of Eq. (\ref{24}) is the first term on the rhs of Fig. 3.
The term $g_{00}^{Nt}$ is the $Nt$ form factor, $({B_t - B_\alpha-{2 \over 3}
q_1^2})^{-1}$ represents the propagation of the  two-particle triton-nucleon
state at a four particle energy $E=-B_\alpha$, the energy for propagation
of the two-particle triton-nucleon state being $(B_t-B_\alpha)$. The term
$({B_d-B_{\alpha}-q_1^2-3q_3^2/4-\vec{q}_1 \cdot \vec{q}_3})^{-1}$
represents the propagation of the three-particle nucleon-nucleon-deuteron
state at a four particle energy $E=-B_\alpha$, the energy for propagation
of the three-particle nucleon-nuclon-deuteron state being $(B_d-B_\alpha)$.
There are two angular momentum-spin coupling coefficients. The one involving
${\bf \hat g}^{Nd}_{{1/2}L_1}$ gives the angular momentum coupling to form
the triton and its coupling to nucleon 1 to give the final zero total angular
momentum of $^4He$. The one involving
${\bf \hat g}^{NN}_{1L_3}$ gives the spin-angular momentum coupling
of nuclons 1 and 2 to form
the deuteron of momentum $-q_3$ and its coupling to the structureless
deuteron 3 to give the final zero total angular
momentum of $^4He$. Finally, there is summation over the internal angular
momenta $L_1$ and $L_3$, and integrations over the internal loop momentum
$\vec q_1$ and angles of $\vec q_3$.

Substituting the values of vertex functions and rearranging
Eq. (\ref{24}), we
obtain the  properly normalized
function $\Lambda^{dd}_{\ell_3} (q_3)$ given by
\begin{equation}
\Lambda_{l_3}^{dd} (q_3) \equiv g_{0l_3}^{dd} (q_3) \sqrt{\frac{2 \pi}
{\mu_{dd}}}  =
\frac {2 C_S^{d} C_S^{t} C_S^{\alpha \rightarrow Nt}}{\pi} \sqrt{
\frac{\mu_{NN} \mu_{Nd} \mu_{Nt}}{\mu_{dd}}}
\int_0^{\infty} dq_1 q_1^2
\frac{{\cal I}_{l_3}(q_3, q_1)}{B_t - B_\alpha - \frac{2}{3} q_1^2}
\label{25} \end{equation}
such that   at the $^4He$ pole it gives the
asymptotic normalization parameters of $ ^{4}He$:
\begin{eqnarray*}
\Lambda^{dd}_{\ell_3} (i\mu_{dd}) =  C^{\alpha \rightarrow dd}_{\ell_3}   .
\label{26} \end{eqnarray*}
In Eq. (\ref{25})
\begin{eqnarray}
{\cal I}_{l_3}(q_3,q_1)& = &
 \int d\Omega_{q_3}d\Omega_{q_1}
\left(({\bf\hat g}_{1L_3}^{NN}(\vec{p}_3) \otimes {{\chi_{1}^d}})^\ast_{l_3}
\otimes Y_{l_3}^*(\hat q_3)\right)_{00} \nonumber \\ & \times  &
\frac {1}{B_d - B_\alpha - q_1^2-
\frac{3}{4} q_3^2- {\vec{q}_1 } \cdot{
\vec{q}_3}} \left( \left({\bf\hat g}_{\frac{1}{2}L_1}^{Nd}({\vec p}_1) \otimes
{\chi_{\frac{1}{2}}^N}\right)_{0}
\otimes Y_{0}(\hat{q}_1) \right)_{00}
\label{27} \end{eqnarray}

 The values of $l_3=0,2$ yield the $S$ and $D$ state
of $ ^{4}He$ respectively.  For evaluating the integral $\cal I$, we
expand the energy propagator in terms of spherical harmonics as below,
\begin{equation}
\frac{1}{B_d-B_{\alpha}-q_1^2-\frac{3}{4}q_3^2-\vec{q}_1 \cdot \vec{q}_3}
= 2{\pi} \sum_{L=0}^\infty {(-1)^L K_L(q_1,q_3) {\sqrt{2L+1}}}
\left(Y_{L}(\hat{q}_1)\times Y_{L}(\hat{q}_3)\right)_{00},
\label{28}  \end{equation}
where
\begin{equation}
K_L(q_1,q_3)=\int_{-1}^1
\frac{P_L(x)\,dx}{B_d-B_{\alpha}-q_1^2-\frac{3}{4}q_3^2-{q}_1\,
{q}_3\,x}.
\label{29} \end{equation}
By using the angular momentum algebra techniques \cite{Rose}, the intrinsic
spin
dependence of the integrand and the part containing the spherical harmonics
 in Eq. (\ref{25}) are
easily separated out and evaluated independent of each other.  Next the same
procedure is adopted to separate the $q_1$ and $q_3$ dependent parts of the
integrand.  After integrating over angles we get the following result
for $\Lambda_{l_3}^{dd}$
\begin{eqnarray}
\Lambda_{l_3}^{dd} (q_3) & = & \frac{C_S^d\,
C_S^t\, C_S^{\alpha \rightarrow Nt}}{\pi} \sqrt{\frac{\mu_{NN}
\, \mu_{Nd}
\, \mu_{Nt}}{\mu_{dd}}}
\int_0^{\infty} dq_1 \frac{q_1^2}{B_t -
B_{\alpha}- {\frac{2}{3}} q_1^2 }
(-1)^{\frac {L_1 +L_3}{2} +1 +l_3 - \beta -\gamma +S_1
-{1\over 2}} \nonumber \\ & \times  &
\sum_{L_1,L_2,\alpha.\beta,\gamma}
\sum_{L=0}^{\infty} K_L (\vec{q}_3, \vec{q}_1)
\left(\frac{\eta_d}{\mu_{NN}^2}\right)^{\frac{L_3}{2}}
\left(\frac{\eta_t}{\mu_{t}^2}\right)^{\frac{L_1}{2}}
\frac{q_1^{\alpha+\beta}\, q_3^{L_1+L_3-\alpha-\beta}
\,2^{\alpha+\beta-L_3+1}}{3^{ \beta}} \nonumber \\ & \times  &
C_{000}^{\beta L \alpha}\, C_{000}^{L_1-\beta L \gamma}
\,C_{000}^{L_3- \alpha l_3\gamma}
\,U(1 L_3 1 l_3;1 L_1) \nonumber \\ & \times  &
U({1\over2}{1\over2}L_1 1 ; 1 S_1)
\,U(\alpha L_3- \alpha L_1 l_3; L_3 \gamma)
\,U(\alpha \gamma \beta L_1-\beta; L_1 L) \nonumber \\ & \times  &
\left[\frac{(2 \beta+1)(2l_3+1)}{(2L+1)(2 \gamma+1)}\right]^{1\over 2}
\left[\frac {(2 L_3+1) ! 2L_1 !}{(2 \alpha+1) ! (2 L_3-2 \alpha)
! (2 \beta+1) ! (2 \L_1- 2 \beta)
!} \right]^{1\over2}.
\label{30} \end{eqnarray}
Here $U(j_1 j_2 j_3 j_4; JK) \equiv (2J+1)(2K+1) W(j_1 j_2 j_3 j_4; JK)$ are
renormalized $6j$ symbols.

As $L_1 = 0$  or $2$ and $L_3 = 0$ or $2$,
the left hand side of the above equation
contains four terms. We retain the three terms linear in
$D$-state and neglect
the term containing a product of $\eta^d$ and $\eta^t$.   [In the limit
$q_3 \rightarrow i \mu$, analytic expressions are easily obtained for $K_L
({q}_1, q_3)$ (relevant $L$ values in the present context being
$L=0,1,2$).]  The asymptotic $D$ state to $S$ state ratio for $^4He$ is
defined by
\begin{equation}
\rho^{\alpha} \equiv \frac {\Lambda_2^{dd}(i \mu_{dd)}}{\Lambda_0^{dd}(i
\mu_{dd})}. \end{equation}
After substituting numerical values of various angular momentum coupling
coefficients  for allowed values of angular momenta in Eq. (\ref{30}), we
evaluate $\rho^\alpha$ as,
\begin{eqnarray}
\rho^\alpha = \mu^2_{dd}
\left(
\frac{\eta^d}{4\mu^2_{NN}} -
\frac{\eta^t}{\mu^2_{nd}}
\right)  -
i\mu_{dd} \frac{F_1}{F_0}
\left(
\frac{\eta^d}{\mu^2_{NN}} -
\frac{4}{3}
\frac{\eta^t}{\mu^2_{Nd}}
\right)
- \frac{F_2}{F_0}
\left(
\frac{\eta^d}{\mu^2_{NN}} -
\frac{4}{9}
\frac{\eta^t}{\mu^2_{Nd}}
\right),
\label{31} \end{eqnarray}
where
\begin{eqnarray*}
F_L  =  \int^\infty_0
\frac{dq_1 q^{L + 2}_1}
{B_t - B_\alpha  - \frac{2}{3} q^2,}
\,K_L(q_1, i \mu_{dd}).
\end{eqnarray*}
Similarly,  the $D^\alpha_2$ parameter of $^4He$ is defined as
\begin{equation}
D^\alpha_2  =  - \lim_{q_3\rightarrow 0}
\frac{g_{02}^{dd} (q_3)}{q^2_3\,
g_{00}^{dd} (q_3) }   \equiv- \lim_{q_3\rightarrow 0}
\frac{\Lambda^{dd}_{2} (q_3)}{\Lambda^{dd}_0 (q_3)\, q^2_3}   .
\label{32}
\end{equation}
The integrals appearing in Eq. (\ref{32}), are preformed analytically
for $q_3\rightarrow 0$ and the result for $D^\alpha_2$ is
\begin{eqnarray*}
D^\alpha_2  =
\left(
\frac{\eta^d}{4\mu^2_{NN}} -
\frac{\eta^t}{\mu^2_{Nd}}
\right) - \frac{1}{6}
\frac{2\mu_{Nt} + \sqrt{B_\alpha - B_d}}
{\mu_{Nt} + \sqrt{B_\alpha - B_d}}
\left(
\frac{\eta^d}{\mu^2_{NN}} - \frac{4}{3}
\frac{\eta^t}{\mu^2_{Nd}}\right)
\end{eqnarray*}
\begin{eqnarray}
+ \frac{2}{15}
\frac{\mu^2_{Nt} + \frac{9}{8}
\mu_{Nt} \sqrt{B_\alpha - B_d} +
\frac{3}{8} \left(
B_\alpha - B_d\right)}
{\left(
\mu_{Nt} + \sqrt{B_\alpha - B_d}\right)^2}
\left(\frac{\eta^d}{\mu^2_{NN}} -
\frac{4}{9} \frac{\eta^t}{\mu^2_{Nd}}\right).
\label{33} \end{eqnarray}
Equations (\ref{31}) and (\ref{33}) are the principal results of the present
study.

\section{NUMERICAL RESULTS}

The numerical results for the $D$ state parameters of $^4He$ based on
Eqs. (\ref{31}) and (\ref{33})  are expected to be more reasonable than that
for $^3H$ of Ref. \cite{Fred1} because of three reasons.
Firstly, the approximate analytical treatment of Ref. \cite{Fred1}
employing the diagramatic equation of Fig. 2 for $^3H$ is  more
approximate than the present treatment  employing  Fig. 3 for $^4He$.
This is because
in the former case the neglect of the spin singlet two nucleon state as an
intermediate state is too drastic;  whereas in the latter case
there are no other competing channels if we permit only exchange
of one nucleon as shown in Fig. 2.  The exchange of two nucleons
is possible but is much less likely and is
usually neglected in the treatment of four nucleon dynamics.\cite{Fonseca}
Secondly, the minimal cluster model we are using
is expected to work better when the nucleus is strongly bound and
the constituents ($^2H$ and
$^3H$) are loosely bound.   As $^4He$ is  strongly bound
this approximation is  more true in $^4He$ than in
$^3H$.  Finally  and most importantly, in the present model
 we are taking the different vertices to be essentially
constants as in Eqs.  (\ref{23}) and (\ref{24}) which
corresponds to taking the vertex form factors unity.   This
reduces the dynamical equations essentially to algebraic
relations between the asymptotic normalization parameters.  In
so doing systematic errors are introduced.  The calculation of
the triton asymptotic $D$ to $S$ ratio $\eta^t$ in Ref. \cite{Fred1}
will have the above error.  But $^4He$ asymptotic $D$ to $S$
ratio $\rho^{\alpha}$ of Eq. (\ref{31}) and $D_2^{\alpha}$ of Eq.
(\ref{33}) are
obtained by dividing two equations of type (\ref{26}) -- one for
$l_3 = 0$ and the other for $l_3 = 2$ -- where exactly identical
approximations are made.  This division is expected to reduce
the above systematic error and Eqs. (\ref{31}) and (\ref{33})
are likely to lead to a more reliable estimate of $^4He$ $D$
state compared to the estimate of $^3H$ $D$ state obtained in
Ref.\cite{Fred1}.

Equation (\ref{31}) or (\ref{33}) yields that  for fixed
$B_t$, $B_d$, $\eta^d$, and $\eta^t$,  $\rho^{\alpha}$ and
$D_2^{\alpha}$ are correlated with $B_\alpha$. Specification of
$B_\alpha$ alone is not enough to determine the $^4He$ $D$ state
parameters. We have established in Refs. \cite{Fred1,Fred2,SX0} that
in a dynamical calculation $\eta^t$ is proportinal to $\eta^d$ for fixed
$B_d$ and $B_t$, from which the theoretical estimate of $\eta^t/\eta^d$
was made. This was relevant because of the uncertainty in the experimental
value of $\eta^d$. If this result is used in Eqs. (\ref{31}) or (\ref{33})
it follows that $\rho^{\alpha}$ and
$D_2^{\alpha}$ are proportional to $\eta^d$ for fixed $B_d$, $B_t$, and
$B_\alpha$.

Next the results of the present calculation using Eqs. (\ref{31})
and (\ref{33}) are presented. In Eqs. (\ref{31}) and (\ref{33}), in actual
numerical calculation both $\eta^d$  and $\eta^t$ are taken to be positive.
The positive sign of $\eta^t$ is consistent with the order of angular momentum
coupling we use in the present study \cite{SX0}.
In Fig. 4 we  plot $\rho^{\alpha}$ versus $B_\alpha$ calculated  using Eq.
(\ref{31}) for different values of $B_t$ and for $\eta^d=0.027$ and
$B^d$ = 2.225 MeV. The value $\eta^d=0.027$ is the average experimental
value reported in Ref. \cite{Eric}. There is a recent experimental
finding: $\eta^d=0.0256$ in Ref. \cite{Rodn}. For the present illustration
we shall, however,
 use $\eta^d=0.027$. Though the final estimate of the asymptotic
$D$ state parameters of $^4He$ will depend on the value of $\eta^d$ employed,
the general conclusions of this paper will not depend on the choice of this
experimental value of $\eta^d$.
The five  lines in this figure correspond to $B_t$ = 7.0,
7.5, 8.0, 8.5, and 9.0 MeV.
The numerical value of $\eta^t$ for a particular $B_t$ is
taken from the correlation in Ref. \cite{Fred1}.   We find that the magnitude
of $\rho^{\alpha}$ increases with the increase of  $B_\alpha$ for a fixed
$B_t$ and with the decrease  of $B_t$ for a fixed $B_\alpha$. This should be
compared with the correlation of $\eta^t$ with $B_t$ in Ref. \cite{Fred1}.
We also calculated $D_2^\alpha$ using Eq. (\ref{33}) for different values of
$B_t$ and $B_\alpha$.

More results of our calculation using Eqs. (\ref{31}) and (\ref{33}) are
exhibited in Table 1.   We employed different values of $B_t$ and
$B_\alpha$.  The value $B_\alpha$ = 28.3 MeV and $B_t$ = 8.48 MeV are the
experimental values.   The other values of $B_\alpha$ and $B_t$ are
considered as they are identical with results of theoretical
calculation of Ref. \cite{Schia}. As the values of the binding energies are
crucial\cite{Fred1,Fred2} for a correct specification of the $D$ state
parameters we decided to consider these binding energies obtained in Ref.
\cite{Schia}.
For example, $B_t$ = 8.15 MeV is the mean of $^3H$ and
$^3He$ binding energies obtained  in Ref.\cite{Schia} with the Urbana
potential. For the same potential  they obtained $B_\alpha = 28.2$ MeV,
and $D_2^{\alpha} = -0.24$ fm$^2$, to be compared with the present
$D_2^{\alpha} = -0.15$ fm$^2$. For the Argonne potential they obtained
mean $B_t$ = 8.04 MeV, $B_\alpha = 27.8$ MeV, and
$D_2^{\alpha} = -0.16$ fm$^2$, to be compared with the present
$D_2^{\alpha} = -0.12$ fm$^2$.
But the large change of $D_2^\alpha$ in Ref. \cite{Schia}
from one case to the other is in contradiction with the present study.
  The first row of Table  1 is  the result of our calculation for
$\rho^{\alpha}$ and $D_2^{\alpha}$ consistent with the correct experimental
 $B_t$ and $B_\alpha$ and using
 $\eta^d = 0.027$, $\eta^t/\eta^d = 1.68$:
\begin{eqnarray}
\rho^{\alpha}&\simeq & -0.14  ,\nonumber \\
D_2^{\alpha} &\simeq & -0.12  fm^2   ,
\label{34} \end{eqnarray}
$$\rho^{\alpha}/\mu^2_{dd} D_2^{\alpha} \simeq  1. $$
In   Ref. \cite{Santos3}  it has been estimated that
$\rho^{\alpha}/(\mu^2_{dd} D_2^{\alpha})$ $\simeq$ 0.9 in agreement to the
present finding.

Next we would like to compare the present result with other
(`experimental') evaluations of these asymptotic parameters. Santos et al.
\cite{Santos1} evaluated $\rho^\alpha$ from an analysis of tensor analyzing
powers for $(d,\alpha)$ reactions on $S$ and $Ar$. They employed a simple
one step transfer mechanism,  plane-wave scattering states, zero-range or
asymptotic bound states. Keeping only the dominant angular momentum states for
the transferred deuteron they predicted $\rho^\alpha = -0.21$.

In another study Santos et al.\cite{Santos2} considered the tensor analyzing
power of reaction $^2H({\vec d},\gamma)^4He$ and concluded that agreement with
experiment could be obtained for $-0.5<\rho^\alpha<-0.4$.

Karp et al.\cite{Karp} studied tensor analyzing powers for
$(d,\alpha)$ reactions on $^{89}Y$. They employed simple shell-model
configurations for the nuclei involved, performed a finite-range DWBA
calculation, and represented the $^4He \rightarrow 2^2H$
overlap by an effective
two-body model. From an analysis of the experimental data the authors
concluded $D_2^\alpha =-0.3 \pm 0.1$ fm$^2$.

Tostevin et al.\cite{Tost1} studied tensor analyzing powers for
$(d,\alpha)$ reactions on $^{40}Ca$. They performed a DWBA calculation
with local energy approximation  and concluded
$D_2^\alpha =-0.31$ fm$^2$ and  $\rho^\alpha = -0.22$. But Tostevin\cite{Tost2}
has later warned that this value of $D_2^\alpha$ may have large error.

Merz et al.\cite{Merz} has performed an analysis in order to make  a
more reliable estimate of these parameters. From a study of the
$^{40}Ca(d,\alpha)^{38}K$ reaction at 20 MeV bombarding energy employing
a full finite-range DWBA calculation they predicted
$D_2^\alpha =-0.19 \pm 0.04$ fm$^2$.

Recently, Piekarewicz and Koonin\cite{Koonin} performed a phenomenological
fit to the experimental data of  the $^2H(d,\gamma)^4He$ reaction
and predicted $\rho^\alpha = -0.4$.
{}From a study of cross section of the same reaction, however,  Weller et al.
\cite{Weller2} predicted $\rho^\alpha = -0.2 \pm 0.05$.

Cosidering the qualitative nature of the present study we find that there is
reasonably good
agreement between the present and other studies. This assures that
we have  correctly included the essential mechanisms of the formation of
the $D$ state.

Unlike in the case of $^3H$,   there are two distinct ingredients
for the formation of the $D$ state of $^4He$: $\eta^t$ and $\eta^d$.  This is
clear from  expressions (\ref{31}) and (\ref{33}). This possibility did not
exist in the case of $^3H$ where $\eta^t$ is determined uniquely by $\eta^d$
apart from the binding energies. In the usual optical  potential  study of
the $D$ state of $^4He$ as in Ref. \cite{Santos3} the dependence of
$\rho^{\alpha}$ on  $\eta^t$ is always neglected. This dependence will be
explicit in a microscopic four-particle treatment of $^4He$. Such microscopic
calculations are welcome in the future for establishing the conclusions of
the present study.

\section{SUMMARY}

We have calculated in a simple model the asymptotic
$D$ state parameters for $^4He$.
The present investigation generalizes the
consideration of universality as presented in Refs. \cite{Fred1} and
\cite{Fred2} for the trinucleon system. The universal trend of the
theoretical calculations on the trinucleon system and the consequent
correlations are generalized here  to the case of the $D$-state
observables of $^4He$. The essential results of our calculation
appear in Eq. (\ref{34}).   We have used a minimal cluster
model in our calculation
where essentially the bound state form factors are neglected, thus
transforming the dynamical equation into an algebraic relation between the
different asymptotic parameters.   Dividing two such equations $-$ one for the
asymptotic $S$ state of $^4He$ and other for the asymptotic $D$ state of
$^4He$ $-$ the estimates of Eq. (\ref{34}) are arrived.   As we have pointed
out in Sec. V,  such a division should reduce the systematic error of the
approximation scheme.   Dynamical calculation using realistic four-body
model should be performed in order to see whether the present estimate
(\ref{34}) is reasonable.   At the same time accurate experimental results
are called for.   We have predicted
correlations between $\rho^{\alpha}$ and $B_\alpha$, and
between $D_2^{\alpha}$ and $B_\alpha$ to be found in actual
dynamical calculations.   Such correlations though appear to be
extremely plausible in view of the calculation of $\eta^t$ of
\cite{Fred1}, can only be verified after performing actual
dynamical calculations.

This work was supported in part by the
Conselho Nacional de Desenvolvimento  Cient\'\i fico e Tecnol\'ogico and
Financiadoras de Estudos e Projetos of Brazil.

\newpage
\begin{center}
{\bf Table 1}
\vskip .5cm
\begin{tabular}{|c|c|c|c|c|c|c|}\hline
&&&&&& \\
$B_\alpha$&$B_t$& $\eta^d$& $\eta^t$& $D_2^{\alpha}$&$\rho^\alpha$&
$\frac{\rho^\alpha}{\mu^2_{dd} D_2^{\alpha}}$\\
(MeV)& (MeV) & & &(fm$^2)$ & & \\
&&&&&& \\ \hline
     &      &       &         &        &     &      \\
28.3 & 8.48 & 0.027 & 0.0454 & -0.12 & -0.14 & 1.07 \\
28.2 & 8.15 & 0.025 & 0.0514 & -0.15 & -0.17 & 0.98\\
27.8 & 8.04 & 0.0266 & 0.0430 & -0.12 & -0.14 & 1.05\\
&&&&&&\\ \hline
\end{tabular}
\end{center}

\vspace*{.5cm}
{\bf Table Caption:} Results for $D$ state parameters of $^4He$
calculated using Eqs. (\ref{31}) and (\ref{33}).

\newpage

\vskip 1cm
{\bf Figure Captions}

1. The coupled Schr\"odinger equation for the formation of the $D$ state
in $^2H$.

2. The coupled Schr\"odinger equation for the formation of the $D$ state
in $^3H$.

3. The present model  for the formation of the $S$ and $D$ states
in $^4He$. In numerical calculation only the first term on the right
hand side of this diagram  is retained.

4. The $\rho^\alpha$ versus $B_\alpha$ correlation for fixed $B_t$ = 7, 7.5,
8, 8.5, and 9 MeV and with $\eta^d$ = 0.027 using Eq. (\ref{31}). The
curves are labelled by the $E_t$ values. The
$\eta^t$ values for each line
are taken from the $\eta^t$ versus $E_t$ correlation of
Ref. \cite{Fred1}.

\end{document}